\documentclass{article}
\makeatletter
\def\ps@headings{%
	\def\@oddhead{\mbox{}\scriptsize\rightmark \hfil \thepage}%
	\def\@evenhead{\scriptsize\thepage \hfil \leftmark\mbox{}}%
	\def\@oddfoot{}%
	\def\@evenfoot{}}
\makeatother 

\PassOptionsToPackage{english}{babel}
\usepackage{apacite}
\usepackage{amsmath}
\usepackage{amssymb}
\usepackage{amsthm}
\usepackage{babel}
\usepackage{bm}
\usepackage{booktabs}
\usepackage[font=small,labelfont=bf]{caption}
\usepackage{emptypage}
\usepackage{epigraph}
\usepackage{fancyhdr}
\usepackage{fancyvrb}
\usepackage{float}
\usepackage{fontenc}
\usepackage{fourier}
\usepackage{geometry}
\usepackage{graphicx}
\usepackage{inputenc}
\usepackage{mathrsfs}
\usepackage{microtype}
\usepackage{multirow}
\usepackage{pifont}
\usepackage{subfigure}
\usepackage{thmtools}
\usepackage{titling}
\usepackage[table]{xcolor}
\usepackage[urlcolor=blue]{hyperref}

\usepackage{mathtools}
\numberwithin{equation}{section}
\numberwithin{figure}{section}
\numberwithin{table}{section}
\DeclareMathOperator*{\argmin}{arg\,min}

\allowdisplaybreaks
\usepackage{tabularx}
\usepackage{algorithm, algorithmicx, algpseudocode}

\newcommand{\mq}{\hspace{.3cm}}

\newcommand{\eqdef}{\stackrel{def}{=}}
\newcommand{\E}[1]{\mathbb{E}\left[#1\right]}
\newcommand{\Esub}[2]{\mathbb{E}_{#1}\left[#2\right]}

\newcommand{\Vsub}[2]{\mathbb{V}\mathrm{ar}_{#1}\left[#2\right]}

\newcommand{\Pb}[1]{\mathbb{P}\left[#1\right]}

\newcommand{\cout}[1]{}
\newtagform{Eq}{[}{]}
\usetagform{Eq}
\newcommand{\vect}[1]{\textrm{vec}\left(#1\right)}
\newcommand{\veps}{\varepsilon}
\newcommand{\nra}{\mathcal N}
\newcommand{\wra}{\mathcal W}
\newcommand{\mcS}{\mathcal S}
\newcommand{\mcK}{\mathcal K}
\newcommand{\bH}{\mathbf H}
\newcommand{\bP}{\mathbf P}
\newcommand{\bu}{\mathbf u}
\newcommand{\bv}{\mathbf v}
\newcommand{\bfy}{\mathbf y}
\newcommand{\bS}{\bm\Sigma}

\usepackage[square,numbers,sort]{natbib}
\title{An Adaptive Learning Approach to Multivariate Time Forecasting in Industrial Processes}
\author{
	Fernando Miguelez$^a$$^b$ 
	\and Josu Doncel$^c$
	\and Maria Dolores Ugarte$^a$$^b$
}
\date {%
	\small
	$^a$ Department of Statistics, Computer Science and Mathematics, Public University of Navarre\%
	$^b$ Institute for Advanced Materials and Mathematics (InaMat$^2$)\\%
	$^c$ Department of Mathematics, University of the Basque Country, UPV/EHU\\%
	\ \\%
	\large\today
}

\begin{document}

\maketitle

\section*{Abstract}
Industrial processes generate a massive amount of monitoring data that can be exploited to uncover hidden time losses in the system. This can be used to enhance the accuracy of maintenance policies and increase the effectiveness of the equipment. In this work, we propose a method for one-step probabilistic multivariate forecasting of time variables involved in a production process. The method is based on an Input-Output Hidden Markov Model (IO-HMM), in which the parameters of interest are the state transition probabilities and the parameters of the observations' joint density. The ultimate goal of the method is to predict operational process times in the near future, which enables the identification of hidden losses and the location of improvement areas in the process. The input stream in the IO-HMM model includes past values of the response variables and other process features, such as calendar variables, that can have an impact on the model's parameters. The discrete part of the IO-HMM models the operational mode of the process. The state transition probabilities are supposed to change over time and are updated using Bayesian principles. The continuous part of the IO-HMM models the joint density of the response variables. The estimate of the continuous model parameters is recursively computed through an adaptive algorithm that also admits a Bayesian interpretation. The adaptive algorithm allows for efficient updating of the current parameter estimates as soon as new information is available. We evaluate the method's performance using a real data set obtained from a company in a particular sector, and the results are compared with a collection of benchmark models.

\medskip

\noindent \textbf{Keywords}: Adaptive parameter estimates; Hidden Markov Model; Industrial processes; Probabilistic prediction.
\section{Introduction}\label{intro}
Machinery and equipment maintenance is the cornerstone of efficient and reliable production processes in industrial settings. With the increasing digitalization and automation of manufacturing lines, and the introduction of cyber-physical control platforms at the shop-floor level, the amount of available data has grown exponentially. This has led to the development of more sophisticated diagnostic and prognostic methodologies to identify and address small inefficiencies or hidden losses. In this context, industrial engineering experts are focusing on a proactive maintenance concept, which involves the early detection and correction of potential issues. Zwetsloot et al.\cite{zwetsloot:2023} 
propose a method for early detection of changes in the frequency of out-of-control events in two signals, and apply it to study the health condition of escalators in some buildings in Hong Kong. The combination of Machine Learning and Deep Learning knowledge with operational data acquisition has led to the development of fault diagnosis methods for equipment in different working conditions. However, these methods typically focus on the degradation of mechanic components of highly specific equipment, but overlook external factors and other possible interactions that could affect the equipment's normal functioning, making them less effective in predicting overall system failures. Some benchmark examples on this matter are discussed by Yang and Zhong\cite{yang:2022}. 
A proactive approach is generally more effective than a reactive one, which only addresses problems after they arise. Statistical methods, advanced analytics tools and machine learning algorithms have enabled the development of predictive maintenance models that try to predict equipment failures, preventing costly downtime and production losses. The advent of continuous data flow in production processes is the basis of several applications that rely on real-time process monitoring, change point detection and the triggering of alerts in case of unusual trends. These methods belong to a process control concept known as Statistical Process Monitoring (SPM), and have proven to be a valuable resource for equipment health management. Woodall and Montgomery\cite{woodall:2014} 
provide a useful overview of techniques within this area. Unfortunately, since they are based on a mostly reactive approach, these methods still fail to predict the behaviour of the process in the near future and to anticipate far enough unplanned long stops caused by major breakdowns or micro stoppages caused by minor faults. This limitation has motivated the exploration of alternative techniques, such as Hidden Markov Models (HMMs). HMMs are flexible and mathematically robust, and have successfully modelled various applications, including speech \citep{rabiner:1989} 
and handwriting  recognition \citep{fischer:2010}, 
electric consumption and generation forecasting \citep{alvarez:2021}, 
and DNA sequences analysis \citep{wojtowicz:2019}. 
Nevertheless, the research community in the field of industrial process engineering agrees on being cautious about the straight utilization of these models due to the natural complexity and variability of industrial process data \citep{afzal:2017}. 

In this article we propose an innovative approach based on Input-Output Hidden Markov Models (IO-HMM) with adaptive learning for probabilistic multivariate forecasting of operational times in industrial processes. The methodology identifies hidden inefficiencies in production by estimating transitions between operational states and modelling the joint distribution of response variables. Through a dynamical parameter update based on the ''Recursive Prediction Error'' (RPE) technique \citep{azimi:2003}, 
the model continuously improves its predictive capacity as new data is incorporated. The method is designed to be implemented in digital industrial management platforms and is applicable to a wide range of manufacturing sectors, while also being flexible enough to be customized to meet the specific needs of each company, including food processing, automotive, pharmaceutical and electronics, particularly in assembly line production, where minimizing downtime and optimizing workflow are critical for efficiency. This approach presents several key advantages for various stakeholders in an industrial setting. For maintenance managers, it enables early detection of equipment failures, facilitating preventive maintenance actions and reducing unplanned downtime. For production engineers, it enhances operational efficiency by identifying bottlenecks and optimizing production times. Industrial data analysts benefit from a robust analytical framework integrating probabilistic modelling with adaptive learning, allowing for better interpretation of process variability. Finally, for plant managers and executives, this approach facilitates data-driven decision-making by offering accurate predictions on equipment availability, performance, and quality.

A similar approach based on HMMs was proposed in Arpaia et al.\citep{arpaia:2020} 
for detecting faulty conditions in fluid machinery. However, the aim of our model is not only the detection of the next operating mode but also the joint prediction of some operational time variables involved in a production process. From these operational times, one can deduce time losses, which reflect process inefficiencies that often remain undetected or overlooked, and some production effectiveness indices, which provide a reliable measure of the current performance of the process.

One of the challenges to overcome when dealing with HMMs is selecting the appropriate number of hidden states. This is especially meaningful in the context of equipment maintenance, since the hidden states are supposed to account for the general condition of the equipment under consideration. In Roblès et al.\cite{robles:2014}, 
authors examined the performance of different HMM topologies using well-known criteria such as the Bayesian Information Criterion, the Shannon Entropy and the Maximum Likelihood among others. The candidate models had different constraints over the transition matrix and different emission probability distributions but all of them were limited to four hidden states. However, this may not be sufficient for real-world applications where multiple intermediate levels may be present due to a variety of factors. Other authors use additional process' signals  to determine the number of hidden states. Baruah and Chinnam\cite{baruah:2005} 
propose an experimental setting for diagnosing physical failure of drill bits and estimating remaining useful life using two highly correlated signals.  
In our approach, we address this issue by letting the data itself to determine the number of hidden states in a stage prior to the HMM modelling, adhering to general guidelines provided in Chinnam and Baruah\cite{chinnam:2009}. 

As mentioned above, industrial processes are generally non-stationary in nature. One way to address this non-stationarity is to allow the parameter estimates to change over time incorporating explanatory variables in the parameter estimation procedure. Afzal and Al-Dabbagh\cite{afzal:2017}  
deal with a multi-signal process by considering an IO-HMM, an extension of the HMM that includes an input stream of variables that affect both the state transitions and the output densities \citep{bengio:1996}. 
In our approach, we adopt an IO-HMM model in which the parameter estimates depend on past values of the operational times and other process features, such as calendar variables (represented by work shifts) and production references. Following the suggestion of Baruah and Chinnam\cite{baruah:2005} 
, the adaptive algorithm mentioned above ensures continuous parameter updating using the latest data. 

The main novelty of this work is the handling of several process signals to identify potential faults in a challenging environment such as production processes. In particular, we focus on the analysis of multiple variables describing the production process carried out by a piece of equipment. Some of the variables are signals that characterize the health condition of the process, and are used to establish the number of hidden states in the Markov chain of the model. Other variables are signals or process features that are considered to affect the response variables, and thus are used as explanatory variables that have an impact on the parameters of interest, i.e., the state transition probabilities and the parameters of the observations' joint density. The explanatory variables are called in the remainder of this work covariates. Further, the adaptive learning algorithm deployed in the continuous part of the IO-HMM is a multivariate extension of the algorithm presented in Alvarez et al.\cite{alvarez:2021} 

The rest of this paper is organised as follows. To provide better context, in Section \ref{time} we define the operational times and indices relevant to this work, and describe the data of the case study that will be presented later. Section \ref{model} outlines the IO-HMM and the methodology for parameter estimation and forecasting of response variables. Section \ref{implementation} details the implementation process. The application to the real case study is introduced in Section \ref{case}. Finally, in Section \ref{conclusion} we discuss the conclusions.
\section{Time losses in industrial processes}\label{time}
In industrial settings, the production process is subject to inefficiencies that eventually assume the form of either output losses or time losses. When represented by time losses, they can be broadly classified into the following categories \citep{muchiri:2008}: 

\medskip

\begin{enumerate}
	\item Stand By Time (\texttt{SBT}): losses due to scheduled stops such as maintenance or cleaning
	\item Down Time (\texttt{DT}): losses due to unexpected stops such as setup adjustments, failures, or supply outages
	\item Performance Losses Time (\texttt{PLT}): losses due to low production speed and micro-stoppages
	\item Quality Losses Time (\texttt{QLT}): losses associated with defective units and rework.
\end{enumerate}

\begin{figure}[htpb]
	\centering
	\includegraphics[width=0.75\linewidth, clip=true, trim=3.51cm 24.57cm 6.34cm 2.3cm]{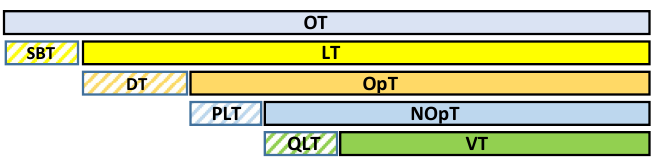}	
	\caption{\normalsize{Production times and time losses classification from an observation period \texttt{OT} \citep{zammori:2011}}}\label{time_losses}
\end{figure}

Note that each of these categories could further be subdivided based on the specific cause of the loss, although such a classification is typically customized according to the particular nature of the process under consideration. By taking the length of an observation period as a reference -hereinafter referred to as Opening Time or \texttt{OT}- one can derive different production times by successively subtracting each time loss, as illustrated in Figure \ref{time_losses} and definitions [\ref{var_defs}].

\begin{minipage}{0.5\textwidth}
	\begin{align*}
		\texttt{OT}-\texttt{SBT}&=\textrm{Loading Time}\ (\texttt{LT})\\
		\texttt{LT}-\texttt{DT}&=\textrm{Operating Time}\ (\texttt{OpT})
	\end{align*}
\end{minipage}%
\begin{minipage}{0.5\textwidth}
	\begin{align}\label{var_defs}
		\texttt{OpT}-\texttt{PLT}&=\textrm{Net Operating Time}\ (\texttt{NOpT})\nonumber\\
		\texttt{NOpT}-\texttt{QLT}&=\textrm{Valuable Time}\ (\texttt{VT})
	\end{align}
\end{minipage}

\bigskip

Moreover, the ratio between the production times can be used to define some well-known effectiveness indicators, which are enumerated in formulae [\ref{kpi_defs}]:

\begin{minipage}{0.5\textwidth}
	\begin{align*}
		\dfrac{\texttt{LT}}{\texttt{OT}}&=\textrm{Loading Rate}\ (\texttt{lo})\\\nonumber\\
		\dfrac{\texttt{OpT}}{\texttt{LT}}&=\textrm{Availability Rate}\ (\texttt{av})
	\end{align*}
\end{minipage}%
\begin{minipage}{0.5\textwidth}
	\begin{align}\label{kpi_defs}
		\dfrac{\texttt{NOpT}}{\texttt{OpT}}&=\textrm{Performance Rate}\ (\texttt{pf})\nonumber\\\nonumber\\
		\dfrac{\texttt{VT}}{\texttt{NOpT}}&=\textrm{Quality Rate}\ (\texttt{qu})
	\end{align}
\end{minipage}

The Overall Equipment Effectiveness (\texttt{OEE}) is a widely-used index that weighs the actual capacity of equipment compared to its optimal capacity, and is defined as the product of the availability, performance and quality rates:
\begin{equation}\label{oee_def1}
	\texttt{oee}=\texttt{av}\times\texttt{pf}\times\texttt{qu}
\end{equation}
The \texttt{OEE} is designed to trace the losses that are directly dependent on the equipment being used, while leaving out other losses that cannot be fixed by rearranging or repairing the equipment. Equivalently, the \texttt{OEE} can also be defined as
\begin{equation*}
	\texttt{oee}=\dfrac{\texttt{LT}-\texttt{DT}-\texttt{PLT}-\texttt{QLT}}{\texttt{LT}}=\dfrac{\texttt{VT}}{\texttt{LT}}
\end{equation*}
or
\begin{equation*}
	\texttt{oee}=\dfrac{\texttt{TU}-\texttt{DU}}{\texttt{ics}\times\texttt{LT}},
\end{equation*}
where \texttt{ics} is the ideal cycle speed (in cycles per time unit; a cycle, or unit, is a produced item), \texttt{TU} is the total number of units and \texttt{DU} the number of defective units. The last definition demonstrates that a $100\%$ value in the \texttt{OEE} is obtained under optimal working conditions, that is, when the process has produced only flawless items at the ideal speed during the scheduled working hours. In Zammori et al.\cite{zammori:2011}, 
the \texttt{OEE} is treated as a random variable and its distribution is used to assess the effectiveness of correction actions implemented in the maintenance strategy. In this study, different time losses are considered as independent beta random variables, but the independency assumption might be taken as unrealistic in real-life processes where, for example, a major failure is often preceded or followed by slower production speeds or by a higher number of rejected units. Our research will also explore whether the dependency between losses leads to a better predictive model by comparing the performances of the multivariate model and the respective univariate models.

To analyse these time losses and develop a predictive model for operational efficiency, we use a dataset collected from a real industrial process. The data comes from a company in a specific industrial sector, where a digital platform is integrated into the manufacturing line to capture and store real-time production data. The dataset comprises 1928 observations collected over four consecutive weeks, and include 35 variables related to the production process. Each observation corresponds to a specific period of operation, with key attributes listed in Table \ref{variables}. A sample of two entries of the dataset is provided in Table \ref{data_table}, which illustrates the structure of the recorded observations and highlights the key variables involved in the analysis.

\begin{table}[htpb]
	\footnotesize\centering
	\caption{Key attributes in the real dataset.}\label{variables}	
	\begin{tabular}{@{\hspace{0.5\tabcolsep}} *3r @{\hspace{0.5\tabcolsep}}}
		\arrayrulecolor{gray}\toprule[1 pt]
		Alias & Variable & Units/format \\ 
		\arrayrulecolor{gray}\toprule[1 pt]
		\multicolumn{3}{l}{\textbf{Production identifiers}}\\
		\texttt{n}      & observation ID       & integer\\
		\texttt{date}   & date                 & yyyy-mm-dd\\
		\texttt{start}  & timestamp            & hh:mm:ss\\
		\texttt{shift}  & workshift            & weekday-shift\\
		\texttt{pr.ord} &  production order ID & integer\\
		\multicolumn{3}{l}{\textbf{Process parameters}}\\
		\texttt{ics}    & ideal unit speed     & units/minute\\
		\texttt{rcs}    & real unit speed (\texttt{TU}/\texttt{LT})& units/minute\\
		\texttt{TU}     & total units          & integer\\
		\texttt{DU}     & defective units      & integer\\
		\texttt{TgU}    & target units (\texttt{OpT}$\times$\texttt{ics}) & real\\
		\texttt{nstops} & number of stops      & integer\\
		\multicolumn{3}{l}{\textbf{Time variables}}\\
		\texttt{OT}     & Opening Time         & minutes\\
		\texttt{SBT}    & Stand By Time        & minutes\\
		\texttt{LT}     & Loading Time         & minutes\\
		\texttt{DT}     & Downtime             & minutes\\
		\texttt{OpT}    & Operating Time       & minutes\\
		\texttt{PLT}    & Performance Losses Time& minutes\\
		\texttt{NOpT}   & Net Operating Time   & minutes\\
		\texttt{QLT}    & Quality Losses Time  & minutes\\
		\texttt{VT}     & Valuable Time        & minutes\\
		\multicolumn{3}{l}{\textbf{Indices}}\\
		\texttt{lo}     & loading rate         &$\in[0,1]$\\
		\texttt{av}     & availability rate    &$\in[0,1]$\\
		\texttt{pf}     & performance rate     &$\in[0,1]$\\
		\texttt{qu}     & quality rate         &$\in[0,1]$\\
		\texttt{oee}    & OEE index            &$\in[0,1]$\\
		\multicolumn{3}{l}{\textbf{Environmental variables}}\\
		\texttt{hum}    & humidity             &\%\\
		\texttt{temp}   & temperature          &\textcelsius\\
		\arrayrulecolor{gray}\bottomrule[1 pt]
	\end{tabular}
\end{table}

\begin{table}[htpb]
	\footnotesize\centering
	\caption{An example of production data extracted from the real dataset.}\label{data_table}	
	\begin{tabular}{@{\hspace{0.5\tabcolsep}} *6r *8r @{\hspace{0.5\tabcolsep}}}
		\arrayrulecolor{gray}\toprule[1 pt]
		\texttt{n} & \texttt{date} & \texttt{start} & \texttt{shift} & \texttt{pr.ord} & \texttt{ics} & \texttt{TU} & \texttt{DU} & \texttt{TgU} & \texttt{OT} & \texttt{SBT} & \texttt{LT} & \texttt{rcs} & \texttt{lo} \\ \arrayrulecolor{gray}\toprule[1 pt]
		66 & 2022-10-10 & 13:50:24 & Mo M & 305 & 1.88 & 13 & 1 & 13.1 & 9.6 & 0 & 9.6 & 1.35 & 1 \\
		67 & 2022-10-10 & 14:00:00 & Mo A & 305 & 1.88 & 13 & 0 & 13.4 & 9.69 & 0 & 9.69 & 1.34 & 1 \\\arrayrulecolor{gray}\bottomrule[1 pt]\\\arrayrulecolor{gray}\toprule[1 pt]
		\texttt{n} & \texttt{DT} & \texttt{OpT} & \texttt{av} & \texttt{PLT} & \texttt{NOpT} & \texttt{pf} & \texttt{QLT} & \texttt{VT} & \texttt{qu} & \texttt{oee} & \texttt{nstops} & \texttt{hum} & \texttt{temp}\\ \arrayrulecolor{gray}\toprule[1 pt]
		66 & 2.62 & 6.98 & 0.73 & 0.05 & 6.93 & 0.99 & 0.53 & 6.4 & 0.92 & 0.67 & 2 & 64.0 & 24.3 \\
		67 & 2.52 & 7.17 & 0.74 & 0.37 & 6.8  & 0.95 & 0    & 6.8 & 1    & 0.7  & 2 & 64.3 & 24.3 \\ \arrayrulecolor{gray}\bottomrule[1 pt]
	\end{tabular}
\end{table}

The motivation behind this research is to develop a probabilistic forecasting model capable of predicting key operational times in the near future. These predictions enable the identification of hidden inefficiencies and the calculation of process performance indicators, which are essential for proactive maintenance strategies, shop-floor decision-making and overall process optimization. However, industrial production dynamics are complex, driven by multiple interdependent factors, and often subject to both systematic patterns and stochastic variability. Therefore, traditional predictive models may struggle to capture these nuances, either because they oversimplify dependencies or fail to adapt to changing conditions. The multivariate IO-HMM provides a flexible framework for modelling the system’s discrete operational modes using hidden states, incorporating explanatory variables to account for external factors affecting production performance, and dynamically adapting to new data through an adaptive learning algorithm that continuously updates model parameters. The real dataset will allow us to assess how effectively this approach achieves these objectives. It will also highlight the model’s strengths while identifying potential areas for improvement, helping us reveal key aspects that require further attention.
\section{The model}\label{model}
\subsection*{Notation}
Roman letters refer to scalar quantities or variables, lowercase bold letters denote vectors and uppercase bold letters denote matrices. Calligraphic letters refer to sets. $\mathbf 1$ denotes a vector of 1's, $\mathbf 0$ a vector or matrix of 0's, $\mathbf I$ is the identity matrix and $\cdot$ denotes the scalar product. Matrix or vector transposition is denoted by the superscript $^T$. 
\subsection{Input-Ouput Hidden Markov Model}\label{hmm}
\begin{figure}[htpb]
	\centering
	\includegraphics[width=\linewidth, clip=true, trim=1.72cm 11.8cm 4.4cm 2.27cm]{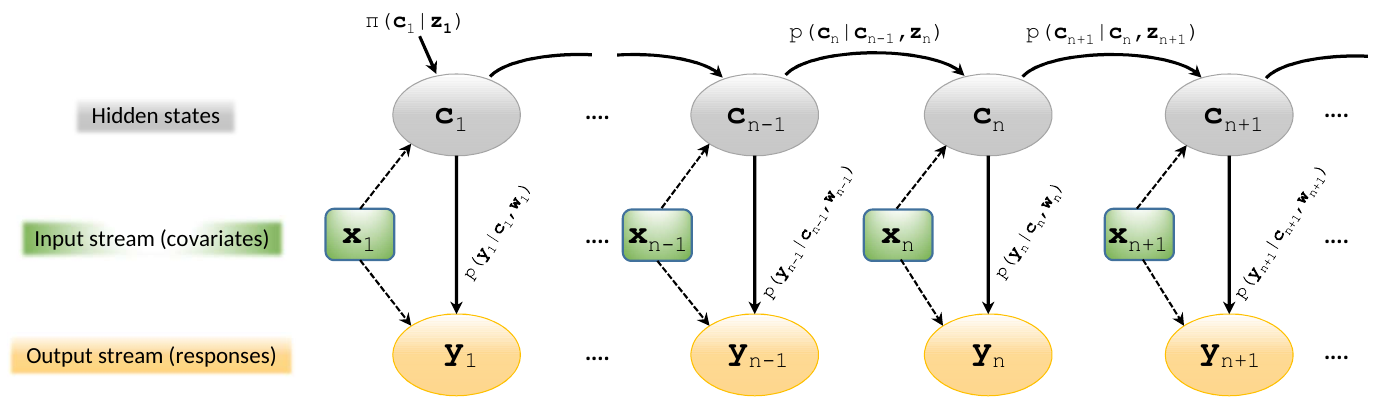}	
	\caption{Diagram of an IO-HMM. Covariates $\mathbf x_n$ affect both discrete and continuous processes. Probabilities in the discrete process $\left\{c_n\right\}_{n\geq1}$ are dependent on covariates $\mathbf z_n$ and probabilities in the continuous process $\left\{\mathbf y_n\right\}_{n\geq1}$ are dependent on covariates $\mathbf w_n$.}\label{io_hmm}
\end{figure}

We model the production process as an IO-HMM, i.e., an HMM with an input stream of covariates. Figure \ref{io_hmm} depicts the diagram of an IO-HMM. The main assumption in HMMs is that the process goes through $K$ hidden states according to a state transition probability distribution. The hidden state of the $n$-th observation period is denoted by $c_n$ and stands for the operational mode of the production process during that period. Each state gives rise to a different probability distribution of the continuous responses in the output stream, that are denoted by $\mathbf y_n$. The decision about the final number of hidden states will be discussed in Section \ref{cluster}. 

The distinctive feature of an IO-HMM is the assumption that the model's parameters are affected by an input stream of covariates. This dependency allows the parameter estimates to change over time, in contrast with the traditional HMM, where the estimates are static after the training phase is completed. The covariates at $n$-th observation period comprehend prior known information about that period and are denoted by $\mathbf x_n$. The covariates affecting the state probabilities in the discrete part of the model will be denoted by $\mathbf z_n$, while those that affect the responses' joint density will be denoted by $\mathbf w_n$, so that $\mathbf x_n=[\mathbf z_n\ \mathbf w_n]$. Both discrete and continuous processes of the model are described in sections \ref{discrete} and \ref{continuous} respectively.

In a simple framework, we can consider that the process's operational mode is exclusively determined by four levels of the \texttt{OEE} index: Optimal (>85\%), Good (60-85\%), Improvable (40-60\%), and Poor (<40\%). The specific operational mode during a given period remains unknown until the observation period ends, when we obtain access to the process information for that specific period, including the \texttt{OEE} score and the response variables $\mathbf y_n$. Based on this \texttt{OEE}, the period is categorized into one of the four levels. Using the prediction error $\mathbf y_n-\hat{\mathbf y}_n$, the last state $c_n$ and the covariates $\mathbf x_{n+1}$, the model parameters are updated. Finally, with the covariates and the latest parameters, the value of the response vector $\hat{\mathbf y}_{n+1}$ is forecasted.

\subsection{The discrete process}\label{discrete}
Assume that the discrete process $\{c_n\}_{n\geq1}$ is a Markov chain with $K$ different states, that is, $c_n\in\left\{1,\ldots,K\right\},\ n\geq1$. The probability distributions for the initial state and the transitions between states are dependent on the covariates $\mathbf z_n$. This vector $\mathbf z_n$ comprises $d$ binary components that describes some features of the period, i.e., $\mathbf z_n\in\left\{0,1\right\}^d,\ n\geq1$.

For a given $\mathbf s\in\left\{0,1\right\}^d$, we assume that the initial probabilities 
$\bm\pi^{(\mathbf s)}=\Pb{c_1|\mathbf z_1=\mathbf s}$ 
and the transition probabilities $\mathbf p_k^{(\mathbf s)}=\Pb{c_n|c_{n-1}=k,\mathbf z_n=\mathbf s},\ k=1,\ldots,K$ follow Dirichlet prior distributions, that is,
\begin{align}
	\bm\pi^{(\mathbf s)}=\left[\pi_1^{(\mathbf s)}\ldots\ \pi_K^{(\mathbf s)}\right]&\sim\textrm{Dirichlet}\left(\mathbf a^{(\mathbf s)}=[a^{(\mathbf s)}_1\ \ldots\ a^{(\mathbf s)}_K]\right)\nonumber\\
	\mathbf p_k^{(\mathbf s)}=\left[p_{k1}^{(\mathbf s)}\ \ldots\  p_{kK}^{(\mathbf s)}\right]&\sim\textrm{Dirichlet}\left(\bm\alpha^{(\mathbf s)}_k=[\alpha^{(\mathbf s)}_{k1}\ \ldots\ \alpha^{(\mathbf s)}_{kK}]\right)\label{dirichlet}.
\end{align}

It is well-known that the parameters of a Dirichlet distribution are recognized as pseudo-counts of the events represented by the random probabilities so that, for example, $a_k^{(\mathbf s)}$ is the pseudo-count of sequences starting  in state $k$ with covariate $\mathbf z_1=\mathbf s$, and $\alpha_{kj}^{(\mathbf s)}$ is the pseudo-count of transitions from state $k$ to state $j$ when the covariates have the value $\mathbf s$.  We will denote by $$a_0^{(\mathbf s)}\eqdef\sum_{k=1}^Ka_k^{(\mathbf s)}\quad \text{ and }\quad \alpha_{k0}^{(\mathbf s)}\eqdef\sum_{j=1}^K\alpha_{kj}^{(\mathbf s)}$$ the concentration parameters of each distribution.

Conditioned to the last observed state, the next unobserved state is a random variable following a categorical distribution (or multinomial with one single trial) with parameters $\bm\pi^{(\mathbf s)}$ if the sequence just begins, or $\mathbf p_k^{(\mathbf s)}$ if the previous observation of the same sequence is in state $k$. Since a prior Dirichlet and a categorical likelihood are conjugate, the posterior distribution for the parameters is also Dirichlet with revised pseudo-counts. Thus, when a new output measurement becomes available it is assigned to the closest state, say $j$, and the relevant pseudo-count is updated by increasing the parameter $a^{(\mathbf s)}_j$ or $\alpha^{(\mathbf s)}_{kj}$ by $1$.
\subsection{The continuous process}\label{continuous}
The continuous process $\{\bfy_n\}_{n\geq1}$ arises from a density function dependent on the state $c_n$ and the covariates $\mathbf w_n$. To model these dependencies, we develop a multivariate extension of the model presented in Alvarez et al.\cite{alvarez:2021}, 
and split the conditional distribution of $\bfy_n|c_n,\mathbf w_n$ into two independent conditional distributions

\begin{subequations}
	\begin{equation}
		\bfy_n|\mathbf w_n \sim N_m
		\left(\bu_n \bH_u, \bS_u\right)\label{cond_dist1}
	\end{equation}
	\begin{equation}
		\bfy_n | c_n \sim N_m(\bv_n \bH_v, \bS_v)\label{cond_dist2},
	\end{equation}
\end{subequations}

\noindent where $m$ is the number of response variables; $\bH_u,\bH_v$ denote coefficient matrices; $\bS_u,\bS_v$ are covariance matrices; and $\bu_n=\left[1\ \mathbf w_n\right]$, $\bv_n=v(c_n)$ are the covariate vectors at the $n$-th step, with $v(\cdot)$ a function of the hidden state. In particular, we propose considering the conditional expectation $v(c_n)=\E{c_n|c_{n-1}}$, which, with the Dirichlet distribution assumptions, simplifies to the Dirichlet parameters in [\ref{dirichlet}] normalized by their concentration parameters, i.e., $\bv_n=\mathbf a/a_0$ or $\bv_n=\bm\alpha_{c_{n-1}}/\alpha_{c_{n-1},0}$.
\subsection{The adaptive algorithm}\label{algorithm}
We explain the adaptive algorithm for the case of the parameters in distribution [\ref{cond_dist1}]; the same procedure is applied to the parameters in [\ref{cond_dist2}]. Let $\lambda\in (0,1]$ be a forgetting factor that accounts for the weight of past observations. As soon as a new sample $\bfy_n$ becomes available, the prior estimators $\bH_{n-1},\bS_{n-1}$ are updated to $\bH_n,\bS_n$ (we omit the model subscripts in the parameters for clarity) through an adaptive algorithm described by the following equations:

\begin{subequations}
	\begin{align}
		\gamma_n &= 1 + \lambda\gamma_{n-1}\label{upd_eq1}\\
		\bH_n  &= \bH_{n-1} + \dfrac{\bP_{n-1}\bu_n^T}{\lambda + \bu_n\bP_{n-1}\bu_n^T}\left(\bfy_n - \bu_n\bH_{n-1}\right)\label{upd_eq2}\\
		\bS_n  &= \bS_{n-1} - \dfrac{1}{\gamma_n}\left[\bS_{n-1} - \dfrac{\lambda\left(\bfy_n - \bu_n\bH_{n-1}\right)^T\left(\bfy_n - \bu_n\bH_{n-1}\right)}{\lambda + \bu_n\bP_{n-1}\bu_n^T}\right]\label{upd_eq3}\\
		\bP_n  &= \dfrac{1}{\lambda}\left(\bP_{n-1} - \dfrac{\bP_{n-1}\bu_n^T\bu_n\bP_{n-1}}{\lambda + \bu_n\bP_{n-1}\bu_n^T}\right)\label{upd_eq4}\\\nonumber	
	\end{align}
\end{subequations}

\noindent initialized with $\bH_0=\bm 0,\ \bS_0=\bm 0,\ \bP_0=\mathbf I$ and $\gamma_0=0$.$\gamma_n$ is the total weight of the sample. Note that if $\lambda=1$ then all the observations have the same weight and $\gamma_n=n$ is the sample size; if $\lambda<1$, past data gradually loses influence over time while recent data has a greater impact. Equations [\ref{upd_eq1}]-[\ref{upd_eq4}] can be obtained by extending to the multivariate case the Maximum-Likelihood-based proof provided in Alvarez et al.\cite{alvarez:2021}, 
Theorem. 1. We refer to Appendix \ref{appendix} for an alternative proof using a Bayesian approach.
\subsection{Forecasting}\label{forecast}
After the training step, each distribution [\ref{cond_dist1}]-[\ref{cond_dist2}] produces a forecast of the responses. Later, these forecasts are combined using a minimum-variance criterion to obtain the final prediction \citep{rocazzella:2022}. 
Once a new observation is available the update-prediction loop begins again. In particular, when the parameters are updated after the $n$-th observation is received we can write

\begin{align*}
	\bfy_{n+1,u} = \bu_{n+1}\bH_u + \bm\varepsilon_{n+1},\qquad&\bm\varepsilon_{n+1}\sim N_m(\bm 0,\bm\Sigma_u)\\
	\bfy_{n+1,v} = \bv_{n+1}\bH_v + \bm\nu_{n+1},\qquad        &\bm\nu_{n+1}\sim N_m(\bm 0,\bm\Sigma_v),\\
\end{align*}
and define the weighted process 

\begin{equation*}\bfy_{n+1} = \bfy_{n+1,u} \mathbf D + \bfy_{n+1,v}(\mathbf I-\mathbf D),\end{equation*}
with $\mathbf D=\textrm{diag}\left(\delta_1,\ldots,\delta_m\right)$ a diagonal weight matrix to be determined. The mean and covariance of this process provide a multivariate forecast of the responses at time $(n+1)$ and an estimate of its accuracy, namely

\begin{subequations}
	\begin{align}
		\Esub{n}{\bfy_{n+1}}\eqdef\hat{\bfy}_{n+1}&=\bu_{n+1}\bH_u\mathbf D + \bv_{n+1}\bH_v(\mathbf I-\mathbf D)\label{pr_eq1}\\
		\Vsub{n}{\bfy_{n+1}}\eqdef\hat{\bS}_{n+1} &=\mathbf D\bS_u\mathbf D + (\mathbf I-\mathbf D)\bS_v(\mathbf I-\mathbf D)\label{pr_eq2},
	\end{align} 	
\end{subequations}
where the subindex $n$ in the expectation and variance operators denotes that they are applied given all the information available at time $n$. We note that finding $\mathbf D$ amounts to obtain separately the optimal weight $\delta_j$ for each response, $j=1,\ldots,m$. This weight\cite{alvarez:2021} 
is given by

\begin{equation}\label{weight}
	\delta_j = \dfrac{\sigma_{v,j}^2}{\sigma_{u,j}^2 + \sigma_{v,j}^2},
\end{equation}
where $\sigma_{u,j}^2$ $(\sigma_{v,j}^2)$ is the $j$-th diagonal element of $\bS_u$ $\left(\bS_v\right)$.
\section{Implementation}\label{implementation}
\subsection{Data Segmentation}\label{data_segm}
In an industrial setting, production processes are typically monitored at fixed intervals, and some time variables, such as \texttt{OT}, \texttt{SBT}, and \texttt{LT} in Figure \ref{time_losses}, are often predetermined and treated as deterministic. This is because planned stops, such as maintenance or cleaning operations, follow predefined schedules, and measurement times are usually established in advance. However, in the dataset used for this study, which was described in Section \ref{time}, these variables are in fact random due to the specific method that employs the company that owns the data capture and management platform. Instead of using fixed measurement intervals, observations are recorded based on some operational events that occur randomly, making the observation times also inherently random, with the exception of one observation that is always recorded at the end of each shift. As a result, variables \texttt{OT}, \texttt{SBT}, and \texttt{LT} fluctuate depending on the data capture timing rather than being predefined, which introduces an additional source of variability in the model. If the process was monitored at predefined times and the variables \texttt{OT}, \texttt{SBT}, and \texttt{LT} were available a priori, the variability of the model would be reduced, and we would expect an improvement in the quality of the predictions.

To structure the data appropriately, each work shift is considered as a separate sequence of observation periods. That is, all observations recorded within a single shift form a continuous sequence in the model, and a new sequence begins when a shift change occurs. This segmentation ensures that the temporal dependencies within each shift are preserved, while taking into consideration the usual equipment adjustments that occur between shift changeovers. At the same time, we try to take advantage of the only measurement time that is known in advance, which is at the end of each shift.

\subsection{Variable selection}\label{select}
Covariate selection is a crucial step in the modelling process, as these variables must be meaningfully related to the response variables. In this context, a covariate is any observable characteristic of the process that potentially affects the dynamics of the target variables and, therefore, its inclusion in the model can enhance prediction accuracy. Some covariates influence the evolution of hidden states, affecting the initial and transition probabilities in the HMM, while others directly impact the joint distribution of output variables.

To ensure that the selected covariates genuinely add value to the model, dimensionality reduction methods such as Principal Component Analysis (PCA) can be employed when the set of available variables is too large. These techniques help identify combinations of variables that capture most of the data’s variability without introducing redundancy. However, in industrial environments, the expertise of the production team is key to identifying relevant variables, so covariate selection can also be also based on operational experience and prior correlation analysis with the response variables. A particular case of covariates includes past values of the response variables $\mathbf y_{n-1},\ldots, \mathbf y_{n-q}$, as they may contain useful information about the future evolution of the process. The optimal number of lags to consider can be determined empirically through cross-validation by comparing the predictive performance of models with different autoregressive orders.

In addition to the covariates used for estimating the model parameters, it is also necessary to select a set of variables that enable the classification of observation periods into different operational states of the process. These classification variables are used in the clustering stage, and must be closely related to process efficiency and reflect its overall performance. Metrics such as availability rate, performance rate, OEE, and other key production indices are suitable candidates for this task. The selection of these variables can also be relied on data analysis techniques, or based on expert knowledge of the production process.

\subsection{Clustering}\label{cluster}
To identify the operational modes of the process, observation periods of the training set are grouped into $K\geq2$ classes using an unsupervised classification technique based on the variables collected in the vector $\mathbf t_n$. Since the choice of classification technique is not the main objective of this work, for simplicity we applied the K-Means method at this stage in our case study, although other clustering approaches can certainly be explored; specifically, dynamic cluster merging and separation \citep{lughofer:2011} techniques are particularly suited for non-stationary environments. Their incorporation into this type of model could enhance our understanding of the process dynamics, and is an interesting line of future research.

The optimal number of classes $K$, which corresponds to the number of hidden states in the Markov model, is not predefined but determined automatically. 
To minimize the need for manual model fitting, we establish this number as the minimum required for a goodness-of-fit (i.e., the between-groups-sum-of-squares divided by the total-sum-of-squares) threshold to be reached, adapting the number of hidden states to the actual complexity of the process and preventing both excessive segmentation and insufficient classification of the data. This will allow the application of the same method in settings with different conditions and equipment. 

Once the classification is complete, each sequence in the training set is segmented into labelled intervals corresponding to the detected operational states. These states are later used to learn the parameters of the discrete part of the model, as described in Section \ref{discrete}, starting from non-informative Jeffreys' priors for the Dirichlet distributions [\ref{dirichlet}], that is, $a_k=\alpha_{jk}=1/2,\ \forall j,k$. 

Let $\mathbf o_k$ be the centroid of the $k$-th class, $\mathbf t_n$ the classification variables, and $c_n$ the class of the $n$-th observation. During the test phase, new observation periods are assigned to the closest centroid, that is  $$c_n=\argmin_{k\in\{1,\ldots,K\}}d(\mathbf o_k, \mathbf t_n),$$ where $d(a,b)$ is a distance function such as Euclidean or Mahalanobis. Alternatively, the labels of the new observations can be obtained through a k-nearest neighbours classification scheme.

\subsection{Pseudo-code}\label{pseudocode}
Let $\mcS$ be the set of values of $\mathbf z_n$ and $\mcK$ the set of hidden states. We define the sets of parameters 

\begin{subequations}
	\begin{align*}
		\bm\Pi    &=\left\{\mathbf{a_s}=[a^{(\mathbf s)}_1\ \ldots\ a^{(\mathbf s)}_K], \mathbf{A_s}=\left(\alpha_{kj}^{(\mathbf s)}\right)_{k,j\in\mcK},\ \mathbf s\in\mcS\right\}\\&\\
		\bm\Psi  &=\left\{\bH_u^{(\mathbf s)},\bS_u^{(\mathbf s)},\bH_v^{(\mathbf s)},\bS_v^{(\mathbf s)},\ \mathbf s\in\mcS\right\}\\&\\
		\bm\Omega&=\left\{\bP_u^{(\mathbf s)},\gamma_u^{(\mathbf s)},\bP_v^{(\mathbf s)},\gamma_v^{(\mathbf s)},\ \mathbf s\in\mcS\right\}
	\end{align*}
\end{subequations}
where
\begin{itemize}
	\item $a_k^{(\mathbf s)}$ is the count of sequences starting in state $k$ with covariates $\mathbf s$,
	\item $\alpha_{kj}^{(\mathbf s)}$ is the count of transitions from state $k$ to state $j$ for observations with covariates $\mathbf s$,
	\item $\bH_{\;\bullet}^{(\mathbf s)}, \bS_{\;\bullet}^{(\mathbf s)}$ are coefficient and covariance matrices respectively,
	\item $\bP_{\;\bullet}^{(\mathbf s)}, \gamma_{\;\bullet}^{(\mathbf s)}$ are state matrices and discount factors respectively.
\end{itemize}

The pseudocode for the learning and forecasting methods is presented in Algorithms \ref{learn_alg} and \ref{pred_alg}.

\begin{algorithm}[htpb]
	\caption{Learning}\label{learn_alg}
	\begin{algorithmic}[1]
		\Require 
		\Statex \mq$\bm\Pi$ \Comment{Dirichlet parameters}
		\Statex \mq$\bm\Psi$ \Comment{model parameters}
		\Statex \mq$\bm\Omega$ \Comment{state parameters}
		\Statex \mq$\lambda_u,\lambda_v$ \Comment{forgetting factors}
		\Statex \mq$\mathbf x_n=\left[\mathbf z_n\ \mathbf w_n\right]$ \Comment{covariates}
		\Statex \mq$c_{n-1}, c_n, \mathbf y_n$ \Comment{state labels, responses}
		\Ensure 
		\Statex $\bm\Pi,\ \bm\Psi,\ \bm\Omega$ \Comment{updated parameters}
		\vspace{0.1cm}
		\For{$n=1,\ldots,N$}
		\State $\mathbf s \gets \mathbf z_n$
		\State $k \gets c_{n-1}$
		\State $\mathbf b \gets \begin{cases}\mathbf a^{(\mathbf s)}&\text{if the sequence begins},\\\bm\alpha^{(\mathbf s)}_k&\text{otherwise}\end{cases}$
		\State $\mathbf u\gets \left[1\ \mathbf w_n\right]$
		\State $\mathbf v\gets \mathbf b/\left(\mathbf b\cdot\mathbf 1\right)$
		\State Update $\gamma_u^{(\mathbf s)},\mathbf H_u^{(\mathbf s)},\bm\Sigma_u^{(\mathbf s)},\mathbf P_u^{(\mathbf s)}$ using equations [\ref{upd_eq1}], [\ref{upd_eq2}], [\ref{upd_eq3}], [\ref{upd_eq4}] respectively
		\State Update $\gamma_v^{(\mathbf s)},\mathbf H_v^{(\mathbf s)},\bm\Sigma_v^{(\mathbf s)},\mathbf P_v^{(\mathbf s)}$ using equations [\ref{upd_eq1}], [\ref{upd_eq2}], [\ref{upd_eq3}], [\ref{upd_eq4}] respectively
		\State $j \gets c_n$
		\If{the sequence begins} 
		\State $a^{(\mathbf s)}_j\gets a^{(\mathbf s)}_j + 1$
		\Else
		\State $\alpha_{kj}^{(\mathbf s)}\gets\alpha_{kj}^{(\mathbf s)} + 1$	
		\EndIf
		\EndFor
	\end{algorithmic}
\end{algorithm}

\begin{algorithm}[htpb]
	\caption{Forecast}\label{pred_alg}
	\begin{algorithmic}[1]
		\Require 
		\Statex \mq$\bm\Psi$ \Comment{model parameters}
		\Statex \mq$\left\{\mathbf o_1,\ldots,\mathbf o_K\right\}$ \Comment{centroids}
		\Statex \mq$\mathbf t_n$ \Comment{classification variables}
		\Statex \mq$\mathbf x_{n+1}=\left[\mathbf z_{n+1}\ \mathbf w_{n+1}\right]$ \Comment{covariates}
		\Ensure 
		\Statex $\hat{\mathbf y}_{n+1},\ \hat{\bm\Sigma}_{n+1}$ \Comment{responses forecast, prediction error}
		\vspace{0.1cm}
		\State $k \gets \argmin_{j\in\{1,\ldots,K\}}d(\mathbf o_j, \mathbf t_n)$
		\State Update $\mathbf o_k$ including $\mathbf t_n$
		\State $\mathbf s \gets \mathbf z_{n+1}$
		\State $\mathbf b \gets \begin{cases}\mathbf a^{(\mathbf s)}&\text{if the sequence begins}\\\bm\alpha^{(\mathbf s)}_k&\text{otherwise}\end{cases}$
		\State $\mathbf u \gets \left[1\ \mathbf w_{n+1}\right]$
		\State $\mathbf v \gets \mathbf b/\left(\mathbf b\cdot\mathbf 1\right)$
		\State Compute $\mathbf D$ using equation [\ref{weight}]
		\State Compute $\hat{\mathbf y}_{n+1}$ using equation [\ref{pr_eq1}]
		\State Compute $\hat{\bm\Sigma}_{n+1}$ using equation [\ref{pr_eq2}]
	\end{algorithmic}
\end{algorithm}

Algorithm \ref{learn_alg} is responsible for updating the model parameters as new data become available. It follows an adaptive learning approach, ensuring that both discrete and continuous parts of the model are refined. The algorithm starts by constructing the necessary feature vectors $\mathbf u, \mathbf v$ for both distribution \ref{cond_dist1}- \ref{cond_dist2} in the continuous part of the model using the covariates and the normalized Dirichlet parameters, as suggested at the end of Section \ref{continuous} (lines 2-6). In the continuous part, the model coefficients and covariance matrices are updated using the adaptive recursive estimation method presented in Section \ref{algorithm} (lines 7, 8). In the discrete part, the algorithm updates the prior state probabilities if a new sequence begins (line 11), or the transition probabilities otherwise (line 13). In any case, it is a simple update of the pseudo-counts that represent the parameters. The process is iterated over all observations (line 1), progressively refining the model's parameters to better reflect the dynamics of the system. 

Algorithm \ref{pred_alg} is in charge for predicting the next values of the response variables based on the most recent data and model estimates. First, the last observation is assigned to a hidden state based on the closest centroid obtained during the clustering step (line 1), and the centroid is updated dynamically (line 2). Next, feature vectors are created as in Algorithm \ref{learn_alg} (lines 3-6), and the model estimates the response variables using both the hidden state-dependent process and the covariates-dependent process. The final forecast is a weighted combination of these two estimates (lines 8, 9), where the weights are determined based on the estimated variance of each process (line 7).

\begin{figure}[htpb]
	\centering
	\includegraphics[width=\linewidth, clip=true, trim=3.34cm 13.25cm 7.4cm 2.2cm]{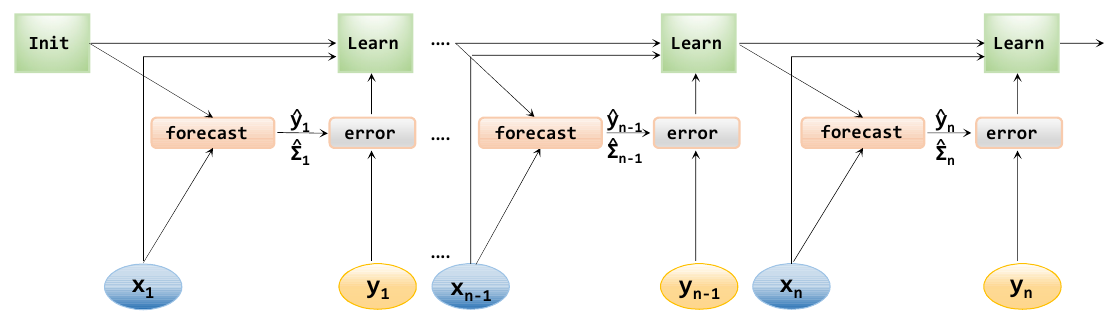}	
	\caption{Block diagram of the adaptive algorithm for the parameter estimates. At the $n$-th step the error in the last prediction, $(\bfy_{n-1}-\hat{\bfy}_{n-1})$, and the new covariates, $\mathbf x_n$, feed the learning algorithm for updating the model parameters. The covariates and the latest parameters are then used to obtain the prediction.}\label{block_diag}
\end{figure}

Figure \ref{block_diag} shows the block diagram of the adaptive algorithm for the model parameter estimates. At each step $n$, the algorithm receives the latest observation $\bfy_{n-1}$ and covariates $\mathbf x_n$, updating the model parameters based on the prediction error $(\bfy_{n-1}-\hat{\bfy}_{n-1})$. These updated parameters are used to generate the next forecast $\hat\bfy_n$ and the covariance matrix $\hat{\bm\Sigma}_n$, renewing the iterative learning-forecasting process.

\subsection{Evaluation}\label{evaluate}
Two standard forecasting metrics are computed for the evaluation task -Mean Absolute Error (MAE) and Root Mean Squared Error (RMSE)- for each response variable. For a test set of L observations, these metrics are defined as

\begin{subequations}
	\begin{align*}
		\text{MAE}=\dfrac{1}{L}\sum\limits_{n=1}^L|y_n-\hat y_n|\\
		\text{RMSE}=\sqrt{\dfrac{1}{L}\sum\limits_{n=1}^L\left(y_n-\hat y_n\right)^2},
	\end{align*}
\end{subequations}
where $y_n$ is the real value and $\hat y_n$ is the forecast. Further, to measure the probabilistic performance of the proposed model, we also compute for each response variable the coverage probability, which is defined as the proportion of real values that fall into the prediction interval:

\begin{equation*}
	covg=\dfrac{1}{L}\sum\limits_{n=1}^L\mathbb I_{\left\{y_n\in\left[\hat y_n\pm1.96\hat\sigma\right]\right\}},
\end{equation*}
where $\mathbb I$ is an indicator function and $\hat\sigma$ is the prediction error. 

We compare the performance of the proposed model with different numbers of response lags, $q=1,\ldots 5$, in the autoregressive component against the following benchmark models:
\begin{itemize}
	\item The \textbf{persistence model}, which uses the last available observation to forecast the next one, that is $\hat{\bfy}_n=\bfy_{n-1}$. This is the baseline model, as all others are expected to produce better predictions.
	\item The \textbf{no-lags model}, which employs adaptive parameter learning without lag responses in the covariates (i.e., $q=0$).
	\item The \textbf{Vector AutoRegressive model with exogenous variables} VARX($q$), $q=1,\ldots 5$. The general form of a VARX($q$) model is \citep{tsay:2013} 
	\begin{equation*}
		\bfy_n=\bm\phi_0+\sum\limits_{j=1}^q\bm\phi_j\bfy_{n-j}+\bm\beta\mathbf g_n+\bm\eta_n,
	\end{equation*}
	where $\bm\phi_j$ are VAR coefficient matrices, $\bm\beta$ is the coefficient matrix for the exogenous variables $\mathbf g_n$ and $\bm\eta_n$ is a sequence of iid random vectors of zero mean and positive-definite covariance matrix. To make a sound comparison, we include in $\mathbf g_n$ the same covariates as in $\mathbf w_n$ without the lagged responses, i.e., $\mathbf w_n=[\mathbf g_n\ \bfy_{n-1}\ \cdots\ \bfy_{n-q}]$. The VARX model does not account for the various operational modes of the process, and it is also a static model, meaning that the parameter estimates are not updated after the training stage. This makes this model adequate to assess the effect of both the inclusion of the discrete process described in Section \ref{discrete} and the adaptive algorithm of Section \ref{algorithm}.
	\item The respective \textbf{univariate models}, which will be used to check whether the multivariate approach takes advantage of the correlation structure between the responses. 
\end{itemize}
	\section{Application to a real case study}\label{case}
In this work, the model has been applied to real data provided by an industrial company. The availability of suitable public datasets for this type of research is very limited, mainly due to confidentiality agreements. Most open datasets related to industrial processes do not provide sufficiently detailed information on operational times, system states, or efficiency indicators, making it challenging to validate the model under realistic conditions. On the other hand, while simulated data could have been an alternative for testing the model in a controlled setting, generating realistic synthetic data for industrial processes is not straightforward. Simulating the dynamics of operational variables, hidden states, and interactions between different process factors requires assumptions that may not faithfully reflect the behaviour of a complex industrial system. For these reasons, the proposed model has been used to predict operational times in the production process of a company operating in a specific industrial sector. Due to confidentiality reasons, the exact industry of this company has been withheld. The data, described in Table \ref{data_table}, is supplied by the technological firm responsible for installing and maintaining the digital platform at the plant. 

To properly feed the presented method, the data has been preprocessed, debugged and arranged using some well-known libraries in \texttt{Python} language \citep{vanrossum:1995} 
such as \texttt{pandas} \citep{mckinney:2010} 
and \texttt{numpy} \citep{harris:2020}. 
The core implementation of the procedure, as described in Section \ref{implementation}, has been developed in \texttt{R} language \citep{r_core:2022} 
, and particularly the VARX models have been fitted using the MTS package \citep{mts:2022}. 

The model's primary objective is to provide probabilistic forecasts of key operational times, thereby allowing production managers to anticipate inefficiencies and improve decision-making.  For instance, if the model predicts an increase in performance losses (PLT) when switching between product types, experienced operators can be reassigned to these transitions to optimize setup times.  Furthermore, forecasts of higher quality-related time losses (QLT) can prompt preemptive quality checks, enabling adjustments to machine settings to minimize defective units.  Finally, if the model identifies patterns of increasing downtime (DT) near the end of long production runs, maintenance teams can schedule preventive maintenance activities before failure occurs.

The time variables \texttt{OpT} and \texttt{NOpT} are selected as response variables due to their strong correlation (0.83), which fully justifies the use of a multivariate model. According to the scheme presented in Figure \ref{time_losses} and the definitions [\ref{var_defs}], [\ref{kpi_defs}] and [\ref{oee_def1}], predicting these variables allows for the computation of time losses due to low production speed and the performance index. Additionally, in scenarios where time measurements are equispaced and scheduled stops are predetermined (i.e., with deterministic \texttt{OT}, \texttt{SBT} and \texttt{LT}), these predictions also enable the estimation of time losses due to unexpected stops and the availability index. 

The covariates in the discrete model ($\mathbf z_n$) include dummy variables that represent work shift levels, affecting the probability distribution of hidden states. The covariates in the continuous model ($\mathbf w_n$) incorporate the same dummy variables along with the ideal unit speed (\texttt{ics}) and two indicators marking the first observation of each shift and the first observation of each production order. Additionally, an autoregressive component is considered by including past response values, $\mathbf y_{n-1},\ldots, \mathbf y_{n-q}$, within $\mathbf w_n$. For the classification step, the most relevant variables related to process health -including availability rate (\texttt{av}), performance rate (\texttt{pf}), overall equipment effectiveness (\texttt{oee}), opening time (\texttt{OT}), real unit speed (\texttt{rcs}) and total units produced (\texttt{TU})- are used as classification criteria. These variables are collected in the vector $\mathbf t_n$.

A careful choice of forgetting factors $\lambda_u, \lambda_v$ is essential to achieving a good trade-off between predictive accuracy and stability. One possible approach, not explored in this study but worth considering, is the use of dynamic forgetting factors\citep{gupta:2018}, where $\lambda_u$ and $\lambda_v$ are adapted over time based on recent process variability. This could allow the model to better balance stability and responsiveness, particularly in non-stationary environments. Our initial experiments indicated that low values for both parameters produce very ill-conditioned state matrices $\mathbf P_n$, leading to unstable and unreliable predictions. This issue is mitigated when the values fall within the range of $[0.9,1]$. From that point on, we found that for $0.9\leq\lambda_u\leq0.92$, the no-lag model consistently outperforms all other models across various metrics, and it remains the best in terms of RMSE and coverage for values $0.93\leq\lambda_u\leq0.95$. It is only when $\lambda_u\geq0.96$ that lagged models begin to show improved performance, though the differences among lag-order models diminish progressively. Additionally, the MAE and RMSE metrics prove to be relatively stable for values of $\lambda_v\geq0.9$, while coverage continues to improve at higher values. Therefore, we select the forgetting factors $\lambda_u=0.99,\ \lambda_v=0.95$ over a grid of values in the interval $[0.9,1]$, although similar nearby values can yield comparable performance. 

\begin{figure}[htpb]
	\centering
	\includegraphics{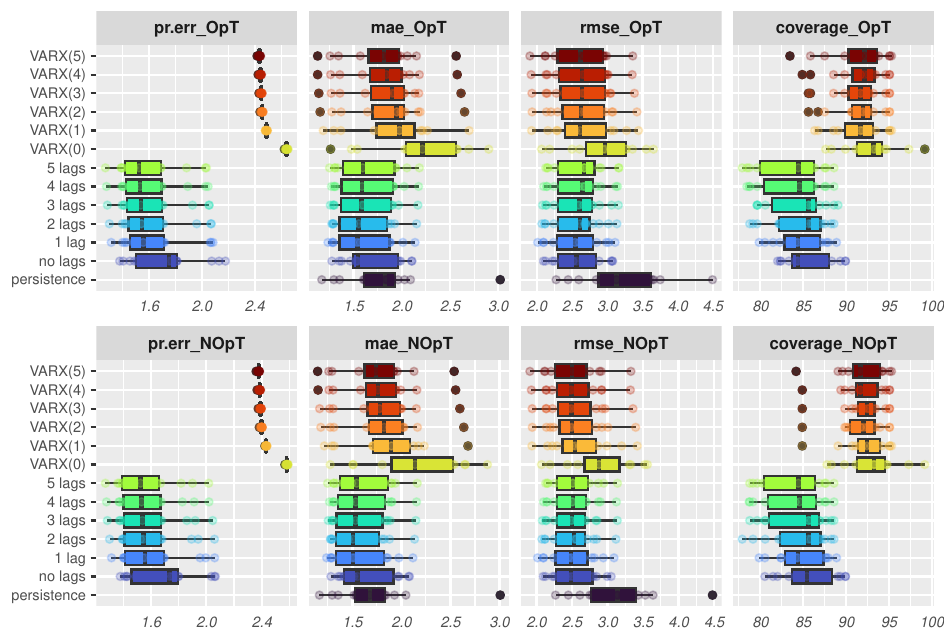}	
	\caption{Boxplot of average prediction error, MAE, RMSE and coverage (by columns) across shifts for each output variable \texttt{OpT} and \texttt{NOpT} (by rows). The persistence model, no-lags model and VARX($q$) models perform worse than the proposed model in terms of MAE, but VARX($q$) models obtain comparable RMSE and better coverage due to the larger prediction error.}\label{boxplot_all}
\end{figure}

Six different models with $q$ responses lags included in $\mathbf w_n$, $q=0,1,2,3,4,5$, are fitted using a Leave-One-Week-Out method, alternatively using three weeks of the dataset in the training step and the fourth week for prediction. MAE, RMSE and coverage are computed separately for each type of shift. We note that the average MAE and RMSE magnitudes are reasonable considering the responses sample quantiles and mean in the dataset, shown in Table \ref{resp_quant}. Figure \ref{boxplot_all} provides an insight into the distribution of the prediction error, MAE, RMSE and coverage across shifts for each output variable in this case study. The box-plot layouts suggest some key points: 
\begin{enumerate}
	\item [(i)] The proposed model outperforms all other models in terms of MAE. Since the MAE measures the mean absolute error, without giving greater weight to larger errors, this result suggests that the combination of an IO-HMM with adaptive learning captures the overall structure of the process better than other models, making it a suitable choice for optimizing processes where average performance is most important.
	
	\item [(ii)] The improvement with respect to the RMSE metric is not as clear, being noticeable only for models with low lag-orders. This points to large errors occurring with a frequency comparable to that of other models, so adjustments should be considered if the objective is to detect uncommon events. 
	
	\item [(iii)] The prediction error of the proposed model is much smaller than the VAR(q) models, resulting in narrower confidence intervals, which also leads to a drop in coverage compared to the VAR(q) models. This suggests that the model underestimates the uncertainty in its predictions, which, linking with the previous point, can lead to large errors when unexpected deviations occur.
\end{enumerate}
As can be observed, the coverage of the proposed model does not reach the nominal value of 95\%, suggesting that it may be underestimating the variance of the prediction error. However, the extent of this underestimation is unclear. While coverage deviates from 95\%, the comparable RMSE also indicates that extreme errors are not necessarily more frequent, which would be expected if uncertainty were severely underestimated. This phenomenon may be caused by several factors, such as limitations in the variance estimation method, a discrepancy between the assumed and actual error distributions, or an insufficient sample size for a reliable estimation. For example, some combinations of work shifts and hidden states may have too few cases to achieve a robust estimation of the variance of the prediction error.

\begin{table}[htpb]
	\small\centering
	\caption{Responses sample mean and quantiles in the 4 weeks dataset.}\label{resp_quant}	
	\begin{tabular}{@{\hspace{0.5\tabcolsep}} *7r @{\hspace{0.5\tabcolsep}}}
		\arrayrulecolor{gray}\toprule[1 pt]
		\texttt{variable}& \texttt{min} & \texttt{Q1} & \texttt{median} & \texttt{mean} & \texttt{Q3} & \texttt{max}\\
		\arrayrulecolor{gray}\toprule[1 pt]
		\texttt{OpT}  & 0 & 6.77 & 9.35 & 8.50 & 10.10  & 19.20 \\
		\texttt{NOpT} & 0 & 6.77 & 9.35 & 8.44 & 10.08  & 19.02 \\
		\arrayrulecolor{gray}\bottomrule[1 pt]
	\end{tabular}
\end{table}

\begin{figure}[htpb]
	\centering
	\includegraphics{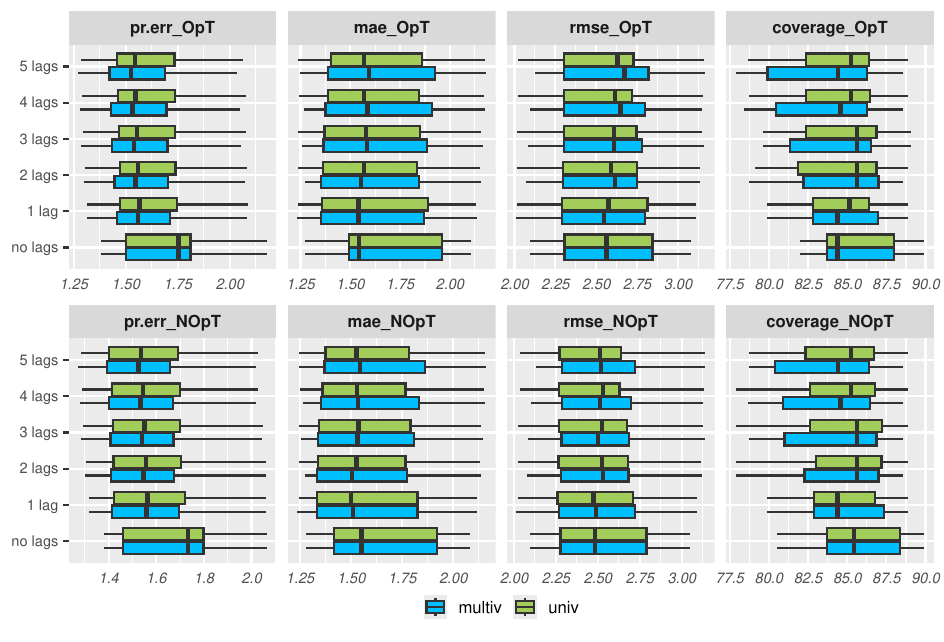}	
	\caption{Comparison between multivariate and univariate models. Despite the high sample correlation between responses (0.83), there is no significant improvement in the performance of the multivariate model.}\label{boxplot_mv_uv}
\end{figure}

Figure \ref{boxplot_mv_uv} presents a comparison between the multivariate model and its respective univariate counterparts. Despite the high correlation between response variables (0.83), which suggests that a multivariate approach should be beneficial, the results do not show a clear advantage of the multivariate model over the univariate ones. One possible explanation lies in the model’s ability to properly estimate uncertainty. As previously discussed, the proposed model exhibits lower coverage than expected, indicating that it may be underestimating the variance of the prediction error. In contrast, univariate models, which estimate each response separately, may be less affected by variance underestimation and thus achieve comparable or even better performance in terms of predictive reliability. Another factor to consider is how well the multivariate structure captures cross-variable dependencies. While a high correlation suggests a strong relationship, it does not necessarily imply that a multivariate model will yield better predictions unless the dependencies are properly exploited. If the model’s formulation does not fully capture the joint dynamics of the responses or the added complexity introduces estimation errors, the expected benefits of the multivariate approach may not materialize. Additionally, limited sample size or misspecification in the variance structure could further hinder its performance.

In summary, while the results of the proposed model applied to the case study seem promising, some areas for improvement have been recognized. Notably, the most critical issues appear to be addressing the underestimation of prediction error variance and improving the exploitation of dependencies between responses.

\section{Conclusions}\label{conclusion}
This study introduces an adaptive learning approach for multivariate time forecasting in industrial processes using an Input-Output Hidden Markov Model (IO-HMM). By enabling probabilistic forecasting of operational times, the model provides a valuable tool for identifying inefficiencies and optimizing production workflows. The results demonstrate that the proposed method can be a helpful alternative against benchmark models, including univariate approaches and Vector AutoRegressive (VARX) models, by leveraging the multivariate nature of the data to capture dependencies between variables more effectively. The integration of an adaptive learning mechanism enables the model to dynamically adjust to new information, ensuring robust performance in non-stationary industrial environments. Through a Bayesian-inspired recursive update process, parameter estimates evolve continuously, making the approach particularly suited for real-time applications.

Beyond its predictive accuracy, the model effectively handles process variability by incorporating exogenous covariates such as production shifts, historical operational data, and other process-specific features. This adaptability, combined with a classification mechanism that automatically determines the number of hidden states, reduces the need for manual adjustments and enhances its applicability across different industrial settings. Designed to be highly scalable, the framework can be implemented in various production environments regardless of differences in equipment or manufacturing processes. Its flexibility allows seamless integration into digital industrial platforms, ensuring long-term usability even as process conditions evolve.

However, despite the theoretical advantages of a multivariate approach, the results indicate that univariate models can be equally competitive or, in some cases, even superior in predictive performance. Several factors could explain this outcome. First, estimating a multivariate model requires learning not only the relationship between each response variable and its covariates but also the dependencies between the responses. If these dependencies are unstable or vary significantly over time, a multivariate approach may not improve prediction accuracy. Moreover, the complexity of estimating a covariance structure can lead to overfitting, particularly when the sample size is not large enough to provide robust parameter estimates. In cases where the correlation between response variables does not provide substantial additional predictive power, a univariate model can achieve similar or better results with fewer parameters and less risk of overfitting. Another potential limitation arises from the way prediction uncertainty is handled: while the multivariate model improves error estimates, it does not always achieve the expected coverage probability. This suggests that its variance structure might not be optimally calibrated, leading to an underestimation of the variance of prediction errors and narrow prediction intervals. Additionally, if the selected covariates already capture most of the information needed for prediction, the added complexity of modelling interdependencies may not translate into meaningful improvements.

The potential applications of this methodology extend beyond the specific case study, offering valuable insights for predictive maintenance by identifying early signs of equipment failure, enabling proactive strategies that minimize unplanned downtime. In production optimization, the model provides accurate forecasts that support real-time decision-making, helping to reduce bottlenecks and improve resource allocation. Its adaptability to real-time industrial monitoring makes it a reliable tool for continuous performance tracking and anomaly detection, further enhancing quality control by identifying patterns linked to production inefficiencies and allowing for early interventions to reduce waste. The approach can also be leveraged for energy and resource management, optimizing power consumption and scheduling by anticipating fluctuations in production efficiency.

This research highlights the value of combining IO-HMMs with adaptive learning techniques to develop a flexible and effective forecasting tool for complex industrial environments. By continuously updating its parameters and integrating multivariate dependencies, the model offers a powerful solution for improving decision-making in industrial processes. Future research could explore extensions to multi-step forecasting, integration with deep learning techniques, and broader scalability improvements to enhance its applicability across diverse sectors, including supply chain optimization and smart manufacturing systems. Additionally, further investigation into the conditions under which multivariate models provide a significant advantage over univariate approaches would help refine their practical use in industrial settings.
\section*{Acknowledgements}
This research was supported in part by the Government of Navarre under Project 0011-1365-2021-000085 and by the Department of Education of the Basque Government through the Consolidated Research Group MATHMODE (IT1456-22).
\section*{Data Availability Statement}
The data that support the findings of this study are available from the corresponding author upon reasonable request. The code to reproduce the results with anonymized data is available at the repository

\begin{center}
	\href{https://github.com/spatialstatisticsupna/MTF-industrial-process}{https://github.com/spatialstatisticsupna/MTF-industrial-process}.
\end{center}
\section*{Disclosure statement}
No potential conflict of interest was reported by the author(s).

\bibliography{references}
\appendix
\section{Appendix}\label{appendix}
\textit{Proof of equations [\ref{upd_eq1}]-[\ref{upd_eq4}]}
\begin{itemize}
\item[a)] Following the directions of \cite{rossi:2012}, pp. 31--34, 
 suppose a $m$-multivariate regression model with $p$ predictor variables
$$\begin{cases}Y_1=&X\beta_1+\veps_1\\
	Y_2=&X\beta_2+\veps_2\\
	&\cdots\\
	Y_m=&X\beta_m+\veps_m\end{cases}$$
with errors correlated across equations. For the $n$-th observation in a random sample of size $N$,
\begin{align*}
\begin{bmatrix}y_{n1}\\\vdots\\y_{nm}\end{bmatrix}&=\begin{bmatrix}\beta_1^T\\\vdots\\\beta_m^T\end{bmatrix}\begin{bmatrix}x_{n1}\\\vdots\\x_{np}\end{bmatrix}+\begin{bmatrix}\veps_{n1}\\\vdots\\\veps_{nm}\end{bmatrix}\\
\underset{\underset{m\times1}{\uparrow}}{Y_n}&=\underset{\underset{m\times p}{\uparrow}}{B^T_{}}\underset{\underset{p\times 1}{\uparrow}}{X_n}+\underset{\underset{m\times1}{\uparrow}}{\bm\veps_n}\qquad\text{ with }\qquad\bm\veps_n\stackrel{iid}{\sim}\nra_m(\bm0,\Lambda),\ n=1,2,\ldots,N
\end{align*}
and gathering all
\begin{align*}
\begin{bmatrix}Y_1^T\\\vdots\\Y_N^T\end{bmatrix}&=\begin{bmatrix}x_{11}&\cdots&x_{1p}\\
	\vdots&\cdots&\vdots\\x_{N1}&\cdots&x_{Np}\end{bmatrix}\begin{bmatrix}\beta_1&\cdots&\beta_m\end{bmatrix}+\begin{bmatrix}\bm\veps_1^T\\\vdots\\\bm\veps_N^T\end{bmatrix}\\
\underset{\underset{N\times m}{\uparrow}}{Y} &= \underset{\underset{N\times p}{\uparrow}}{X}\underset{\underset{p\times m}{\uparrow}}{B}+\underset{\underset{N\times m}{\uparrow}}{E}
\end{align*}
For some $\Lambda_0\in\mathcal M_{m\times m},\ N_0\in\mathbb N,\ \beta_0\in\mathcal M_{mp\times1},\ V_0\in\mathcal M_{p\times p}$, the natural conjugate priors for the parameters in the multivariate regression model can be taken as
\begin{align*}
	\Pb{\Lambda,B}&=\Pb{B|\Lambda}\Pb{\Lambda}\\
	\Lambda&\sim\wra^{-1}(N_0\Lambda_0,m,N_0+m+1)\\
	\beta|\Lambda&\sim\nra_{mp}\left(\beta_0,\Lambda\otimes V_0^{-1}\right)
\end{align*}
where $\wra^{-1}$ denotes a inverted Wishart distribution, $\otimes$ is the Kronecker product and $\beta=\vect{B}$ (vectorization). The prior means are $\E{B|\Lambda}=B_0$ and $\E{\Lambda}=\Lambda_0$.

Given these priors and the random sample, it is well known that the posterior joint density for the parameters can be decomposed into the product of the following densities
\begin{align*}
	\Lambda|Y,X&\sim\wra^{-1}\left(N_0\Lambda_0+N\tilde S,m,N_0+N+m+1\right)\\
	\beta|\Lambda,Y,X&\sim\nra_{mp}\left(\tilde\beta,\Lambda\otimes\left(X^TX+V_0\right)^{-1}\right)
\end{align*}
where
\begin{subequations}
	\begin{align}
		\tilde\beta&=\vect{\tilde B}\nonumber\\
		\tilde B&=B_0+\left(X^TX+V_0\right)^{-1}X^T\left(Y-XB_0\right)\nonumber\\
		N\tilde S&=\left(Y-X\tilde B\right)^T\left(Y-X\tilde B\right)+\left(\tilde B-B_0\right)^TV_0\left(\tilde B-B_0\right)\label{NStilde}.
	\end{align}
\end{subequations}
Hence, the posterior means are
\begin{subequations}
	\begin{align}
		\E{B|\Lambda}&=\tilde B=B_0+\left(X^TX+V_0\right)^{-1}X^T\left(Y-XB_0\right)\label{post_mean_B}\\
		\E{\Lambda|Y}&=(N_0\underset{\underset{\text{prior mean}}{\uparrow}}{\Lambda_0}+N\tilde S)/(N_0+N)\nonumber\\
		&=\Lambda_0-\dfrac{N\left(\Lambda_0-\tilde S\right)}{N_0+N}\label{post_mean_var}.
	\end{align}
\end{subequations}

\item[b)] To make the above results fit in with our case let us define
\begin{align}
	Y\               &=\ \mathbf y_n\ \text{ (new responses, row vector $1\times m$)}\nonumber\\
	X\               &=\ \mathbf u_n\ \text{ (new predictors, row vector $1\times p$)}\nonumber\\
	\tilde B\        &=\ \mathbf H_n\ \text{ (posterior mean, $p\times m$)}\nonumber\\
	B_0\             &=\ \mathbf H_{n-1}\ \text{ (prior mean, $p\times m$)}\nonumber\\
	V_0\             &=\ \lambda\mathbf P_{n-1}^{-1}\ \text{ (prior state matrix, $p\times p$)}\nonumber\\
	(X^TX+V_0)^{-1}\ &=\mathbf P_n\ \text{ (posterior state matrix, $p\times p$)}\nonumber\\
	\gamma_n\        &=1+\lambda+\cdots\lambda^{n-1}\ \text{(scalar)}\nonumber
\end{align}

Equation [\ref{upd_eq1}] is obvious given that $\gamma_n=1+\lambda\left(1+\lambda+\cdots\lambda^{n-2}\right)=1+\lambda\gamma_{n-1}$. From the above definition of the posterior state matrix $\mathbf P_n$ and using the Sherman-Morrison formula\footnote{$(A+uv^T)^{-1}=A^{-1}-\dfrac{A^{-1}uv^TA^{-1}}{1+v^TA^{-1}u}$},
\begin{align*}
	\mathbf P_n &= \left(X^TX+V_0\right)^{-1}\\
	&= \left(\mathbf u_n^T\mathbf u_n + \lambda\mathbf P_{n-1}^{-1}\right)^{-1}\\
	&=\left(\lambda\mathbf P_{n-1}^{-1}\right)^{-1}-\dfrac{\left(\lambda\mathbf P_{n-1}^{-1}\right)^{-1}\mathbf u_n^T\mathbf u_n\left(\lambda\mathbf P_{n-1}^{-1}\right)^{-1}}{1+\mathbf u_n\left(\lambda\mathbf P_{n-1}^{-1}\right)^{-1}\mathbf u_n^T}\\
	&=\dfrac{1}{\lambda}\left(\mathbf P_{n-1}-\dfrac{\mathbf P_{n-1}\mathbf u_n^T\mathbf u_n\mathbf P_{n-1}}{\lambda+\mathbf u_n\mathbf P_{n-1}\mathbf u_n^T}\right)
\end{align*}
we obtain equation [\ref{upd_eq4}]. For the posterior mean $\tilde B=\mathbf H_n$, using expression [\ref{post_mean_B}] we have
\begin{align*}
	\mathbf H_n=\tilde B&= B_0+\left(X^TX+V_0\right)^{-1}X^T\left(Y-XB_0\right)\\
	&=\mathbf H_{n-1}+\left(\mathbf u_n^T\mathbf u_n+\lambda\mathbf P_{n-1}^{-1}\right)^{-1}\mathbf u_n^T\left(\mathbf y_n-\mathbf u_n\mathbf H_{n-1}\right)\nonumber\\
	&=\mathbf H_{n-1}+\mathbf P_n\mathbf u_n^T\left(\mathbf y_n-\mathbf u_n\mathbf H_{n-1}\right)\nonumber\\
	&=\mathbf H_{n-1}+\dfrac{1}{\lambda}\left(\mathbf P_{n-1}-\dfrac{\mathbf P_{n-1}\mathbf u_n^T\mathbf u_n\mathbf P_{n-1}}{\lambda+\mathbf u_n\mathbf P_{n-1}\mathbf u_n^T}\right)\mathbf u_n^T\left(\mathbf y_n-\mathbf u_n\mathbf H_{n-1}\right)\nonumber\\
	&=\mathbf H_{n-1}+\dfrac{1}{\lambda}\dfrac{\mathbf P_{n-1}\mathbf u_n^T\left(\lambda+\mathbf u_n\mathbf P_{n-1}\mathbf u_n^T\right)-\mathbf P_{n-1}\mathbf u_n^T\mathbf u_n\mathbf P_{n-1}\mathbf u_n^T}{\lambda+\mathbf u_n\mathbf P_{n-1}\mathbf u_n^T}\left(\mathbf y_n-\mathbf u_n\mathbf H_{n-1}\right)\nonumber\\
	&=\mathbf H_{n-1}+\dfrac{\mathbf P_{n-1}\mathbf u_n^T}{\lambda+\mathbf u_n\mathbf P_{n-1}\mathbf u_n^T}\left(\mathbf y_n-\mathbf u_n\mathbf H_{n-1}\right),
\end{align*}
yielding equation [\ref{upd_eq2}]. On the other hand, from the definition [\ref{NStilde}] we get
\begin{align*}
	N\tilde S&=\left(Y-X\tilde B\right)^T\left(Y-X\tilde B\right)+\left(\tilde B-B_0\right)^TV_0\left(\tilde B-B_0\right)\\
	&=\left(\mathbf y_n-\mathbf u_n\mathbf H_n\right)^T\left(\mathbf y_n-\mathbf u_n\mathbf H_n\right)+\left(\mathbf H_n-\mathbf H_{n-1}\right)^T\left(\lambda\mathbf P_{n-1}^{-1}\right)\left(\mathbf H_n-\mathbf H_{n-1}\right)\\
	&=\left[\mathbf y_n-\mathbf u_n\left(\mathbf H_{n-1}+\dfrac{\mathbf P_{n-1}\mathbf u_n^T}{\lambda+\mathbf u_n\mathbf P_{n-1}\mathbf u_n^T}\left(\mathbf y_n-\mathbf u_n\mathbf H_{n-1}\right)\right)\right]^T\left[\mathbf y_n-\mathbf u_n\left(\mathbf H_{n-1}+\dfrac{\mathbf P_{n-1}\mathbf u_n^T}{\lambda+\mathbf u_n\mathbf P_{n-1}\mathbf u_n^T}\left(\mathbf y_n-\mathbf u_n\mathbf H_{n-1}\right)\right)\right]+\\
	&\phantom{=}\ \left[\dfrac{\mathbf P_{n-1}\mathbf u_n^T}{\lambda+\mathbf u_n\mathbf P_{n-1}\mathbf u_n^T}\left(\mathbf y_n-\mathbf u_n\mathbf H_{n-1}\right)\right]^T\left(\lambda\mathbf P_{n-1}^{-1}\right)\left[\dfrac{\mathbf P_{n-1}\mathbf u_n^T}{\lambda+\mathbf u_n\mathbf P_{n-1}\mathbf u_n^T}\left(\mathbf y_n-\mathbf u_n\mathbf H_{n-1}\right)\right]\\
	&=\left[\left(\mathbf y_n-\mathbf u_n\mathbf H_{n-1}\right)-\dfrac{\mathbf u_n\mathbf P_{n-1}\mathbf u_n^T}{\lambda+\mathbf u_n\mathbf P_{n-1}\mathbf u_n^T}\left(\mathbf y_n-\mathbf u_n\mathbf H_{n-1}\right)\right]^T\left[\left(\mathbf y_n-\mathbf u_n\mathbf H_{n-1}\right)-\dfrac{\mathbf u_n\mathbf P_{n-1}\mathbf u_n^T}{\lambda+\mathbf u_n\mathbf P_{n-1}\mathbf u_n^T}\left(\mathbf y_n-\mathbf u_n\mathbf H_{n-1}\right)\right]+\\
	&\phantom{=}\ \lambda\dfrac{\mathbf u_n\mathbf P_{n-1}\mathbf u_n^T}{\left(\lambda+\mathbf u_n\mathbf P_{n-1}\mathbf u_n^T\right)^2}\left(\mathbf y_n-\mathbf u_n\mathbf H_{n-1}\right)^T\left(\mathbf y_n-\mathbf u_n\mathbf H_{n-1}\right)\\
	&=\left[\dfrac{\lambda\left(\mathbf y_n-\mathbf u_n\mathbf H_{n-1}\right)}{\lambda+\mathbf u_n\mathbf P_{n-1}\mathbf u_n^T}\right]^T\left[\dfrac{\lambda\left(\mathbf y_n-\mathbf u_n\mathbf H_{n-1}\right)}{\lambda+\mathbf u_n\mathbf P_{n-1}\mathbf u_n^T}\right]+\lambda\dfrac{\mathbf u_n\mathbf P_{n-1}\mathbf u_n^T}{\left(\lambda+\mathbf u_n\mathbf P_{n-1}\mathbf u_n^T\right)^2}\left(\mathbf y_n-\mathbf u_n\mathbf H_{n-1}\right)^T\left(\mathbf y_n-\mathbf u_n\mathbf H_{n-1}\right)\\
	&=\dfrac{\lambda\left(\mathbf y_n-\mathbf u_n\mathbf H_{n-1}\right)^T\left(\mathbf y_n-\mathbf u_n\mathbf H_{n-1}\right)}{\lambda+\mathbf u_n\mathbf P_{n-1}\mathbf u_n^T}.
\end{align*}

Now, by choosing
\begin{align*}
	N_0\ &=\ \lambda\gamma_{n-1}\\
	N\ &=\ 1\\
	\Lambda_0\ &= \bm\Sigma_{n-1}\ \text{ (prior mean variance, $m\times m$)}\\
	\E{\Lambda|Y}\ &=\ \bm\Sigma_n\quad \text{ (posterior mean variance, $m\times m$)}
\end{align*}
and using expression [\ref{post_mean_var}]
\begin{align*}
	\bm\Sigma_n=\E{\Lambda|Y}\ &= \Lambda_0-\dfrac{N\left(\Lambda_0-\tilde S\right)}{N_0+N}\\
	&= \bm\Sigma_{n-1}-\dfrac{1}{1+\lambda\gamma_{n-1}}\left[\bm\Sigma_{n-1}-\dfrac{\lambda\left(\mathbf y_n-\mathbf u_n\mathbf H_{n-1}\right)^T\left(\mathbf y_n-\mathbf u_n\mathbf H_{n-1}\right)}{\lambda+\mathbf u_n\mathbf P_{n-1}\mathbf u_n^T}\right]\\
	&= \bm\Sigma_{n-1}-\dfrac{1}{\gamma_n}\left[\bm\Sigma_{n-1}-\dfrac{\lambda\left(\mathbf y_n-\mathbf u_n\mathbf H_{n-1}\right)^T\left(\mathbf y_n-\mathbf u_n\mathbf H_{n-1}\right)}{\lambda+\mathbf u_n\mathbf P_{n-1}\mathbf u_n^T}\right]
\end{align*}
follows equation [\ref{upd_eq3}].
\end{itemize}
\end{document}